\documentclass[pdflatex,sn-mathphys-num]{sn-jnl}% Math and Physical Sciences Numbered Reference Style
\usepackage{graphicx}%
\usepackage{multirow}%
\usepackage{amsmath,amssymb,amsfonts}%
\usepackage{amsthm}%
\usepackage{mathrsfs}%
\usepackage[title]{appendix}%
\usepackage{xcolor}%
\usepackage{textcomp}%
\usepackage{manyfoot}%
\usepackage{booktabs}%
\usepackage{algorithm}%
\usepackage{algorithmicx}%
\usepackage{algpseudocode}%
\usepackage{listings}%
\usepackage{lineno}
\theoremstyle{thmstyleone}%
\theoremstyle{thmstyletwo}%

\theoremstyle{thmstylethree}%

\makeatletter
\renewcommand\@fnsymbol[1]{\ifcase#1\or a\or b\or c\or d\or e\or f\or g\or h\or i\else\@ctrerr\fi}
\makeatother

\begin{document}

%\title[Article Title]{Design and development of optical detectors\\for the BUTTON-30 experiment}

\title[Article Title]{Design and development of optical modules\\for the BUTTON-30 detector}

%=============================================================%
%======================modified 4 Dec author list =================%

\author*[1]{D.~S.~Bhattacharya}\email{dbhatta2@ed.ac.uk}
\author[2]{J.~Bae}
\author[3]{M.~Bergevin}
\author[4]{J.~Boissevain}
\author[5]{S.~Boyd}
\author[6]{K.~Bridges}
\author[7]{L.~Capponi}
\author[6]{J.~Coleman}
\author[8]{D.~Costanzo}
\author[9]{T.~Cunniffe}
\author[3]{S.~A.~Dazeley}
\author[10]{M.~V.~Diwan}
\author[3]{S.~R.~Durham}
\author[1]{E.~Ellingwood}
\author[11]{A.~Enqvist}
\author[8]{T.~Gamble}
\author[10]{S.~Gokhale}
\author[6]{J.~Gooding}
\author[2]{C.~Graham}
\author[11]{E.~Gunger}
\author[12]{W.~Hopkins}
\author[2]{I.~Jovanovic}
\author[13,14]{T.~Kaptanoglu}
\author[8]{E.~Kneale}
\author[13,14]{L.~Lebanowski}
\author[9]{K.~Lester}
\author[3]{V.~A.~Li}
\author[~~8]{M.~Malek \footnote{Present communication: \href{mailto:matthew@banterbots.ai}{matthew@banterbots.ai}}}
\author[4]{C.~Mauger}
\author[6]{N.~McCauley}
\author[6]{C.~Metelko}
\author[7]{R.~Mills}
\author[6]{A.~Morgan}
\author[1]{F.~Muheim,}
\author[9]{A.~Murphy}
\author[1]{M.~Needham}
\author[2]{K.~Ogren}
\author[13,14]{G.~D.~Orebi~Gann}
\author[1]{K.~Y.~Oyulmaz}
\author[9]{S.~M.~Paling}
\author[15]{A.~F.~Papatyi}
\author[~~19]{A.~Petts\footnote{Visiting professor at: Department of Physics, Durham University, DH1~3LE, UK}}
\author[9]{G.~Pinkney}
\author[9]{J.~Puputti}
\author[16]{S.~Quillin}
\author[5]{B.~Richards}
\author[10]{R.~Rosero}
\author[~~8]{A.~Scarff \footnote{Present address: \url{http://www.tubr.tech}}}
\author[~~6]{Y.~Schnellbach,\footnote{Present address: Science for Nuclear Diplomacy, Technische Universität Darmstadt, Schlossgartenstr.~7, 64289 Darmstadt, Germany}}
\author[9]{P.~R.~Scovell}
\author[12]{B.~Seitz}
\author[8]{L.~Sexton}
\author[1]{O.~Shea}
\author[~~1]{G.~D.~Smith\footnote{Present Address: Health Physics, Gartnavel Royal Hospital, 1055 Great Western Road, Glasgow~G12~0XH, UK}}
\author[17]{R.~Svoboda}
\author[5]{D.~Swinnock}
\author[6]{A.~Tarrant}
\author[12]{F.~Thomson}
\author[6]{J.~N.~Tinsley}
\author[9]{C.~Toth}
\author[1]{A.~Us\'{o}n}
\author[18]{M.~Vagins}
\author[1]{J.~Webster}
\author[1]{S.~Woodford}
\author[10]{G.~Yang}
\author[10]{M.~Yeh}
\author[5]{E.~Zhemchugov} %

% Affiliations relabeled sequentially by first appearance
\affil[1]{School of Physics and Astronomy, University of Edinburgh, Edinburgh, EH9~3FD, UK}
\affil[2]{Department of Nuclear Engineering and Radiological Sciences, University of Michigan, Ann Arbor, MI~48109, USA}
\affil[3]{Lawrence Livermore National Laboratory, Livermore, CA~94550, USA}
\affil[4]{Department of Physics and Astronomy, University of Pennsylvania, Philadelphia, PA~19104, USA}
\affil[5]{Department of Physics, University of Warwick, Gibbet Hill Road, Coventry, CV4~7AL, UK}
\affil[6]{Department of Physics, University of Liverpool, Merseyside, L69~7ZE, UK}
\affil[7]{United Kingdom National Nuclear Laboratory, Cumbria, CA14~3YQ, UK}
\affil[8]{School of Mathematical and Physical Sciences, University of Sheffield, Sheffield, S10~2TN, UK}
\affil[9]{STFC, Boulby Underground Laboratory, Boulby Mine, Redcar-and-Cleveland, TS13~4UZ, UK}
\affil[10]{Brookhaven National Laboratory, Upton, NY~11973, USA}
\affil[11]{Department of Materials Science \& Engineering, Nuclear Engineering Program, University of Florida, Gainesville, FL~32611-6400, USA}
\affil[12]{School of Physics and Astronomy, University of Glasgow, Glasgow, G12~8QQ, UK}
\affil[13]{University of California, Berkeley, Berkeley, CA, USA}
\affil[14]{Lawrence Berkeley National Laboratory, Berkeley, CA, USA}
\affil[15]{Pacific Northwest National Laboratory, 902 Battelle Boulevard, Richland, WA~99352, USA}
\affil[16]{AWE Aldermaston, Reading, Berkshire, RG7~4PR, UK}
\affil[17]{University of California at Davis, Department of Physics and Astronomy, Davis, CA~95616, USA}
\affil[18]{Department of Physics and Astronomy, University of California, Irvine, CA~92697-4575, USA}
\affil[19]{EDF Energy UK, Hartlepool, TS25~2BZ, UK}

%==================================author list ends==============================%
%%================================================================================%

\abstract{BUTTON-30 is a neutrino detector demonstrator located in the STFC Boulby underground facility in the north-east of England. The main goal of the project is to deploy and test the performance of the gadolinium-loaded water-based liquid scintillator for neutrino detection in an underground environment. This will pave the way for a future large-volume neutrino observatory that can also perform remote monitoring of nuclear reactors for nonproliferation.
This paper describes the design and construction of the watertight optical modules of the experiment.}



\maketitle

\section{Introduction}
\label{sec:intro}
The Boulby Underground Technology Testbed for Observing Neutrinos (BUTTON-30) is a technology demonstrator installed in the Boulby Underground Laboratory (BUL) in the north-east of England \cite{buttonpaper}. The detector is a right cylindrical stainless steel tank with an outer diameter of $3.6 \,$m and a height of $3.2 \,$m that will be filled with 30-tonnes of water-based liquid scintillators loaded with gadolinium (WbLs). Located at a depth of 1070\,m in a polyhalite mine, BUL \cite{Murphy:2012zz} provides an ideal low background environment to study neutrinos and to search for dark matter. The cosmic muon background is suppressed by approximately six orders of magnitude compared to surface level \cite{Robinson:2003zj} and the radioactive background levels from the surrounding rock are low \cite{Malczewski:2013lqy}. 

 BUTTON-30 will study the performance of novel hybrid fill media such as Water-based Liquid Scintillator (WbLS) \cite{Bignell_2015, Xiang_2024, Ascencio-Sosa_2024, WbLS-attenuation-scattering} and gadolinium (Gd) loading \cite{MARTI2020163549, ABE2022166248, ABE2024169480, PhysRevApplied.18.034059}. The use of these novel media have the potential to be transformative for neutrino detection. The optical detector system consists of ninety-six 10-inch Hamamatsu R7081-100 photomultiplier tubes (PMTs) with low radioactivity glass \cite{R7081datasheet}.  The electrical characterisation of a sample of these PMTs, prior to encapsulation, is described in detail in~\cite{Akindele:2023ixz}. To allow operation in WbLs, and other novel fill media, the PMTs are encapsulated in custom-built watertight acrylic housings.
In this paper, the design, assembly and quality assurance of the optical modules leading up to their installation at Boulby is described. The paper is laid out as follows. First, the design of the housing is described. Next, the optical characterisation and pressure qualification are presented. Following this, the assembly and quality assurance procedures are discussed.

\section{Design of the optical module}
\label{sec:Housing}
\begin{figure}
\centering
\includegraphics[scale=0.90]{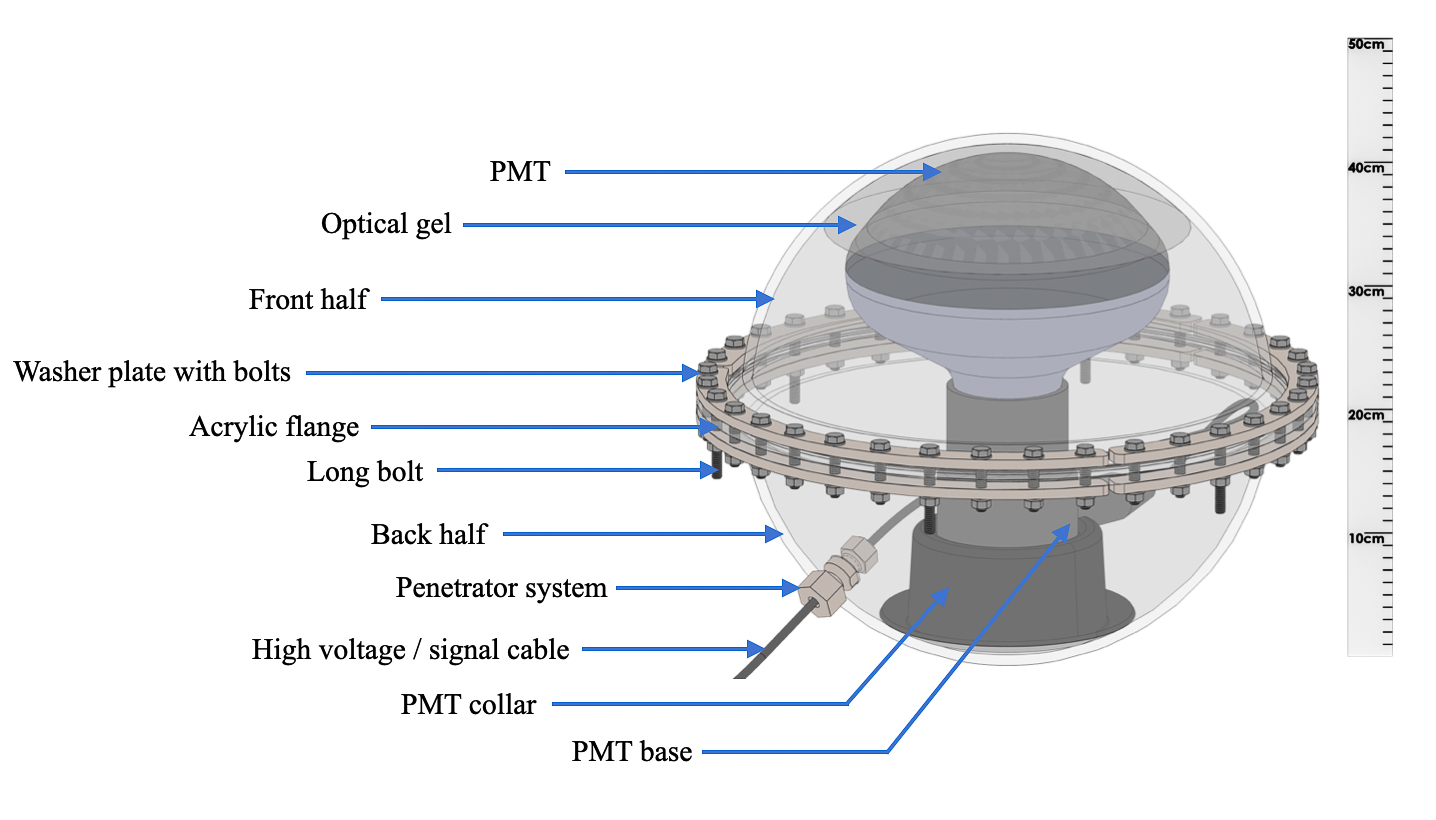}
\caption{\label{fig:CAD-optical-module}Schematic illustration of the acrylic optical housing with an encapsulated 10-inch Hamamatsu R7081-100 PMT. The main components are labelled, including the hemispherical housing halves, flange with O-ring, washer plates, penetrator assembly, and optical coupling gel. This design ensures watertight operation with various fill media. The interior of the back half of an actual housing is painted black.}
%\label{housing-design}
\end{figure}

 Hamamatsu supplied the PMTs with waterproof bases that are compatible with operation in Gd-loaded water. Soak tests have shown that the bases react chemically with WbLS. To allow BUTTON-30 to run with WbLS, each PMT is encapsulated in a spherical acrylic housing, with a design inspired by the IceCube Digital Optical Modules (DOMs) \cite{IceCube:2016zyt, IceCube:Erratum}.  Each housing consists of two acrylic hemispheres of radius 48.6\,cm with a flat flange, as shown in Figure~\ref{fig:CAD-optical-module}. These were produced using a blow moulding technique by a Glasgow-based company, ICLTech ~\cite{ILCTech}. The measured thickness of the hemisphere varies from $3\,$mm at the apex of the dome to $6\,$mm at the flange. Standard acrylic is used for the back half, while the front half uses acrylic that is transparent for ultraviolet (UV) wavelengths. A Viton O-ring sits within a groove in the flange. Two stainless steel washer plates compress the O-ring to give a watertight seal. 

%Optical housing CAD %%%%%%%%%%% Fig1 %%%%%%%%%%%%%
 
 The PMT is optically coupled to the acrylic housing using two-component silicone gel. During the design phase the properties of several gels were evaluated and 
Techsil RTV27905 was selected. This gel is easy to procure, inexpensive and has good adhesive properties. It has good transmission characteristics in the UV region (Section~\ref{sec:OpticalProperties}) and well-matched refractive index of 1.405. A volume of 1.4\,L of gel is sufficient to cover the photocathode region of the PMT. The gel and a plastic collar bonded to the rear of the housing mechanically support the PMT.
 
 High-voltage (HV) and signal return from the PMT are provided by a 25\,m coaxial cable integrated into the PMT base. The cable exits the housing via a watertight assembly referred to as the penetrator system. This consists of a body piece, two Viton O-rings, a ferrule, and two nuts. Tightening the outer nut compresses the ferrule against one of the O-rings, ensuring a watertight seal around the cable. A silica desiccator bag attached to the PMT base controls the humidity level within the housing. The inner surface of the rear of the housing is painted black with acrylic paint to block the reflection of stray-light. The chosen paint, RS Matt-Black Aerosol, reduces light transmission for wavelengths near $400\,\textrm{nm}$ by $99\, \%$. 
 
Care was taken in the selection of materials in contact with the fill media to prevent corrosion or leaching of detector material into WbLS and Gd-doped media \cite{Xiang_2024, MARTI2020163549}. Marine-grade, 316 stainless steel is used for the washer plates, bolts, and parts of the penetrator system. All steel parts were pickled in HF/Nitric acid (ASTM A380) and electropolished to specification ISO 15730. The passivation of a sample of the washer plates was verified using ferroxyl testing and electronic testing by the manufacturer. Soak tests of the acrylic, O-ring material, and PMT cables were carried out to ensure compatibility with WbLS operation. 

%Having established the structural design, we next characterised the optical properties of the housing and coupling gel to confirm compatibility with PMT sensitivity.

\section{Optical characterisation}
\label{sec:OpticalProperties}
\begin{figure}
\centering
\includegraphics[scale=0.6]{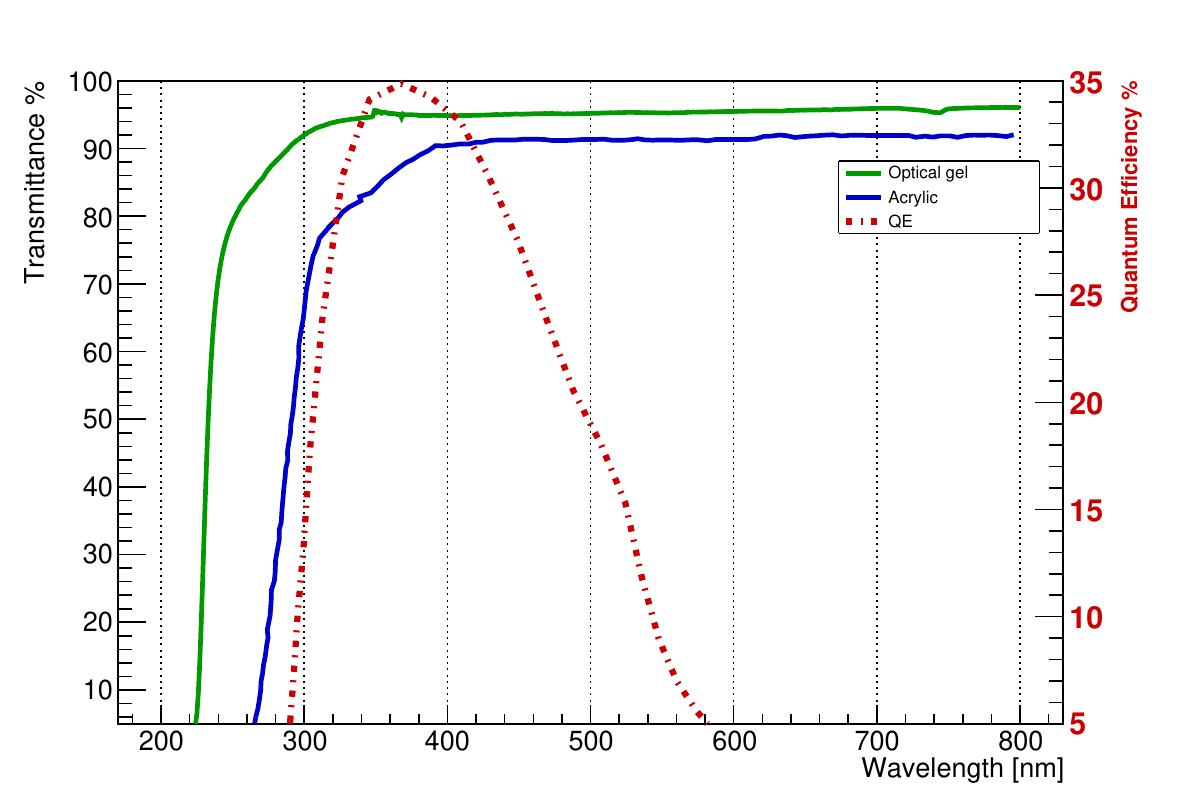}
\caption{\label{fig:light-transparency}Measured transmittance of the RTV27905 optical gel (200–800\,nm) and the selected UV-transparent acrylic, compared with the Hamamatsu R7081-100 photocathode quantum efficiency curve \cite{R7081datasheet}. The transparency of both materials aligns well with the PMT spectral response, ensuring efficient photon collection.}
%\label{Transmittance-QE}
\end{figure}

To match the quantum efficiency (QE) of the PMT photocathode, both the optical gel and the front half of the housing need to be transparent to wavelengths down to $\sim$300\,nm. Several UV-transparent acrylic samples were optically characterised and tested for compatibility with WbLS. Acrylic provided by ICLTech (Glasgow) was selected based on its good transmission characteristics and proximity to the vendor. The transparency of the selected acrylic, shown in Figure~\ref{fig:light-transparency}, is well aligned with the QE of the PMT. The transparency of the optical gel was measured across the 200–800 nm wavelength range with an Agilent UV–Vis spectrophotometer (Figure~\ref{fig:light-transparency}), confirming high transmission in the sensitive wavelength region of the PMT photocathode. The UV–Vis transmission measurements have an estimated systematic uncertainty of $\pm 2\%$, dominated by sample alignment. %Based on the optical measurements, RTV27905 gel was selected, as it demonstrates satisfactory performance at a reasonable cost. 
The overall optical transparency of the module compared to a bare PMT at a wavelength of 400 nm, is approximately 88\% (Figure~\ref{fig:light-transparency}). 

%Gel-Acrylic-QE%%%%%%%%%%%% Fig %%%%%%%%%%%%%%%%%%

\section{Pressure qualification}
\label{sec:PressureQualification}
The housings must be watertight and withstand the hydrostatic overpressure of 3\,m of water ($\sim$0.3\,bar). The pressure tolerance of the housing was evaluated both through direct pressure tests in a water vessel and also finite-element simulations using COMSOL Multiphysics~\cite{COMSOL}. The simulation incorporates surface profile measurements from prototype housings but does not account for micro-cracks in the acrylic, which may reduce the effective strength.

%COMSOL-module-OLD(3)%%%%%%%%%%%% Fig3 %%%%%%%%%%%%%%%
\begin{figure}
\includegraphics[width=0.50\linewidth]{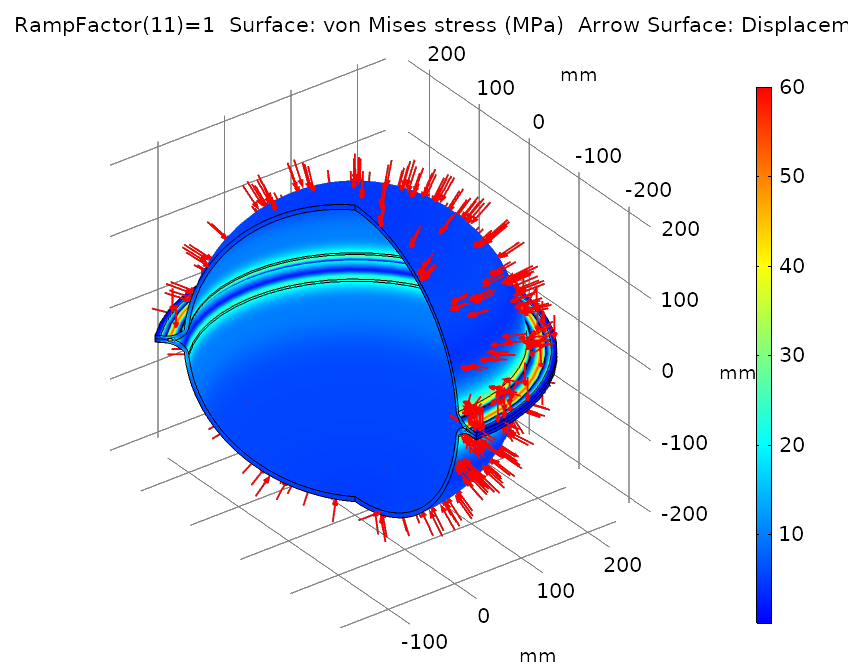}
\includegraphics[width=0.50\linewidth]{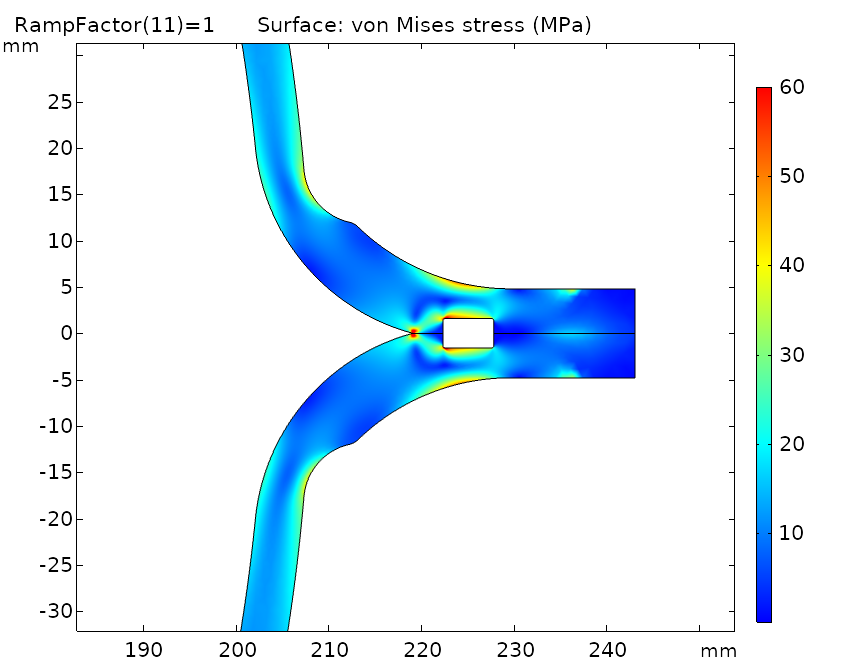}
\caption{ \label{fig:6-COMSOL-old-3D} Finite-element analysis (COMSOL Multiphysics) of a thermoformed acrylic housing under 3\,bar pressure. Left: stress distribution across the hemisphere. Right: zoomed region near the dome-to-flange transition, where maximum stress ($\sim$60\,MPa) is observed due to abrupt thinning of the thermoformed geometry.}
\end{figure}

%COMSOL-module-OLD(4)%%%%%%%%%%%%%% Fig4 %%%%%%%%%%%%%%%
\begin{figure}
\includegraphics[width=0.50\linewidth]{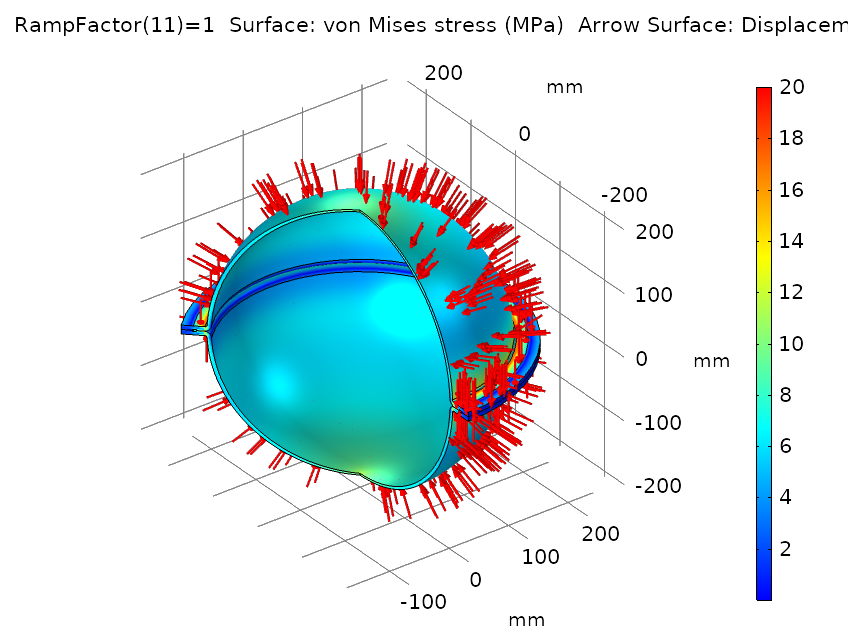}
\includegraphics[width=0.50\linewidth]{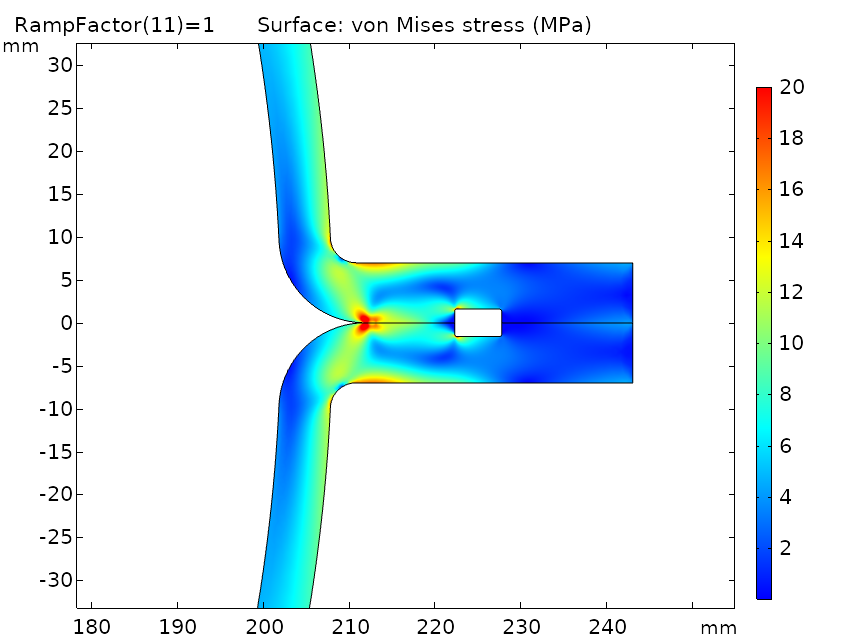}
\caption{ \label{fig:6-COMSOL-new-3D}Finite-element analysis (COMSOL Multiphysics) of the blow-moulded acrylic housing. Left: stress distribution across the hemisphere. Right: region near the flange where stress concentrations occur. The blow-moulded design exhibits reduced maximum stress ($\sim$20\,MPa) and improved uniformity compared to the thermoformed version (Figure~\ref{fig:6-COMSOL-old-3D}).}
\end{figure}

The first prototypes were produced by vacuum thermoforming of acrylic. This technique resulted in variations in thickness and flatness, particularly in the transition region between the dome and the flange, where stresses are concentrated. Under 3\,bar applied pressure, the three-dimensional COMSOL model predicts a maximum stress of $\sim$60\,MPa in this region due to abrupt thinning (Figure~\ref{fig:6-COMSOL-old-3D}). Three prototype thermoformed modules were tested in a high-pressure vessel. All were able to sustain pressures up to 3\,bar. One module was kept at 2\,bar for 12 hours without signs of water ingress (verified using humidity paper placed inside the housing). Destructive implosion of one thermoformed module was observed at 3.2\,bar overpressure, consistent with expectations.

Although the pressure strength of the thermoformed housings exceeds the requirements of BUTTON-30 by an order of magnitude, the prototype housings were fragile. In particular, the flange was susceptible to cracking during handling, which resulted in a low production yield. To address this, ICLTech switched to a blow-moulding process for the final production. This change resulted in a thicker flange region with greater uniformity, significantly improving the durability and handling characteristics.
The wall thickness for the blow-moulded design gradually decreases to approximately $3\,$mm near the apex of the hemisphere. Despite this, the COMSOL simulation (Figure~\ref{fig:6-COMSOL-new-3D}) shows no significant increase in stress at the apex. In addition, the maximum stress around the flange is reduced to $\sim$20\,MPa as seen in Figure~\ref{fig:6-COMSOL-new-3D}. Due to time constraints and the unavailability of the pressure test chamber, destructive tests on the new blow-moulded housings could not be carried out. Nevertheless, based on the performance of the previous iteration and the improved handling characteristics, we are confident that the new housings meet BUTTON’s requirements.

%assembly-steps(5)%%%%%%%%%%% Fig %%%%%%%%%%%%%%%%%%
\begin{figure}
\centering
\includegraphics[scale=0.5]{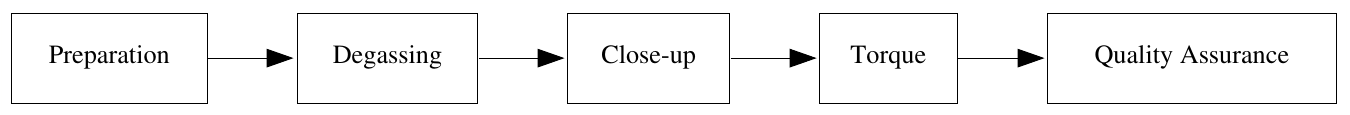}
\caption{\label{fig:steps} Workflow summary of the optical module assembly procedure,  The steps include quality assurance of components, preparation of housing and penetrator, degassing of the optical gel, precise positioning of the PMT, closure of housing halves, controlled torqueing of bolts, and final verification checks.}
%\label{Transmittance-QE}
\end{figure}

\section{Assembly procedure}
\label{sec:AssemblyProcedure}
The assembly method for the optical modules was inspired by the procedure developed for IceCube DOMs~\cite{IceCube:2016zyt}. A well-defined procedure,  with detailed steps and checklists at each stage, is critical to ensure successful production of the optical modules with low failure rate. Although the objectives of the BUTTON-30 experiment did not require assembly in a cleanroom environment, care was taken throughout the production process to maintain a clean working area, and modules were handled with gloves to avoid contamination. At each stage of production, the modules were wiped to minimise residual contamination from sources such as dust, acrylic flecks, or optical gel. The production procedure is summarised in Figure~\ref{fig:steps} and discussed in the following subsections. The assembly of ninety-nine modules was completed in approximately four months, with only a handful of housings requiring rework. Despite the care in defining the procedures, the most common issues encountered were minor variations in operator practice, such as torque application.

\subsection{Preparation}
\label{sec:prep}
The first step of the procedure was the assembly of the back half of the module. The rear hemisphere was painted black, taking care to ensure there were no paint flecks on the flange or the outer parts of the housing. Once the paint had cured, the penetrator system was assembled. The PMT cable was then thoroughly cleaned and pulled through the penetrator. 

\subsection{Degassing}
\label{sec:degas}
%Rig-Holder(6)%%%%%%%%%%%%%%%%% Fig %%%%%%%%%%%%%%%%%%%
\begin{figure}
\includegraphics[width=0.49 \linewidth]{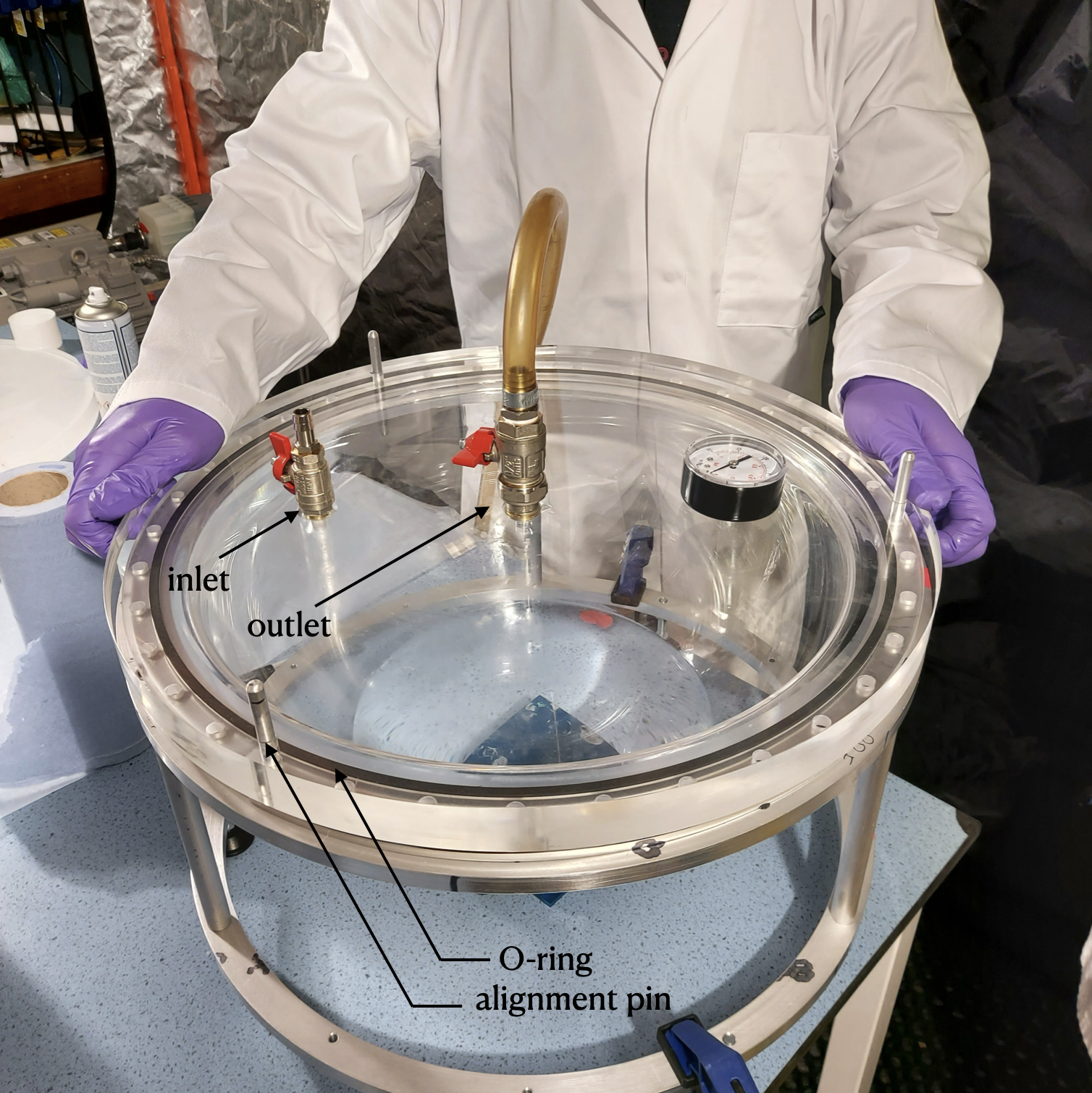}
\includegraphics[width=0.49 \linewidth]{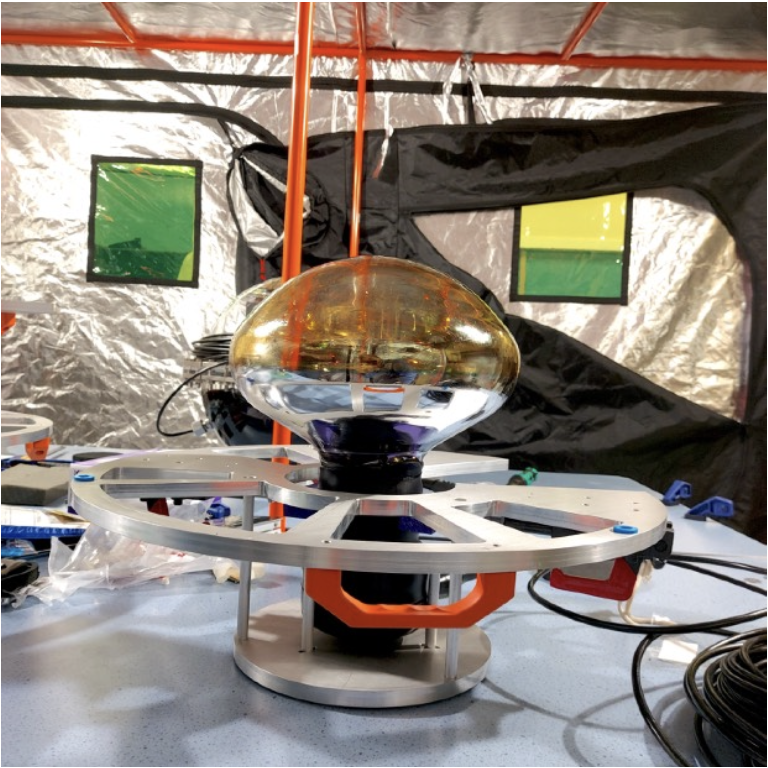}
\caption{\label{fig:10-rig-mixed-gel} Assembly process of the front hemisphere. Left: mixed RTV27905 optical gel being poured into the acrylic dome prior to degassing. Right: 10-inch PMT secured vertically in the alignment holder, ensuring correct positioning during gel curing.}
\end{figure}

A mounting rig, shown in Figure~\ref{fig:10-rig-mixed-gel}, held the front half with the concave surface facing upward. The three short alignment pins ensured consistent positioning of housing components throughout assembly. Subsequently, the O-ring was inserted into the groove of the flange. A holder, illustrated in Figure~\ref{fig:10-rig-mixed-gel} (right), was used to mount the PMT vertically. The holder employed two clamps, positioned on opposite sides to securely attach the PMT base. Before tightening the clamps, care was taken to ensure that the PMT base was vertically aligned.

%Before-After_Degassing(7)%%%%%%%%%%%% Fig %%%%%%%%%%%%%
\begin{figure}
\includegraphics[width=0.49\linewidth]{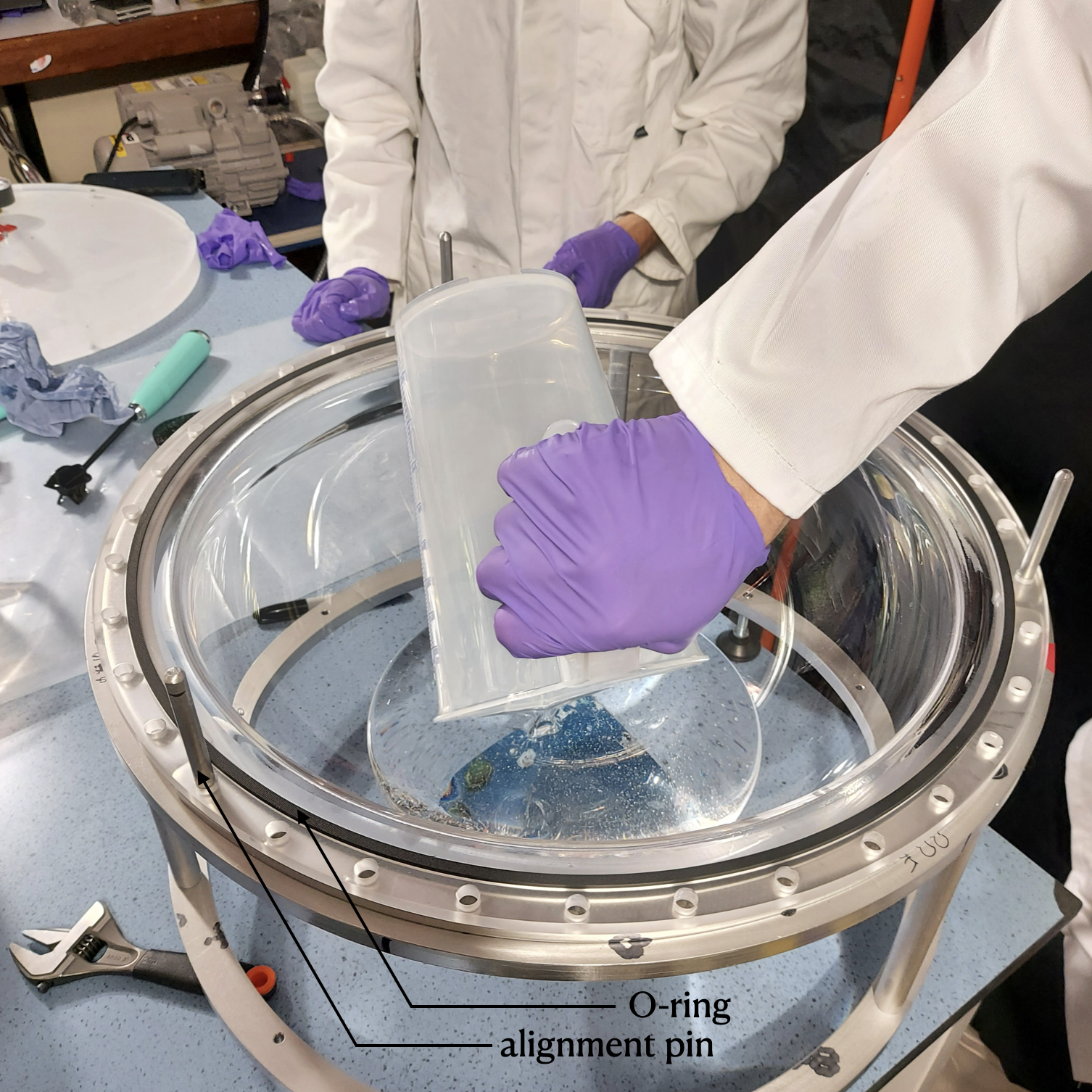}
\includegraphics[width=0.49\linewidth]{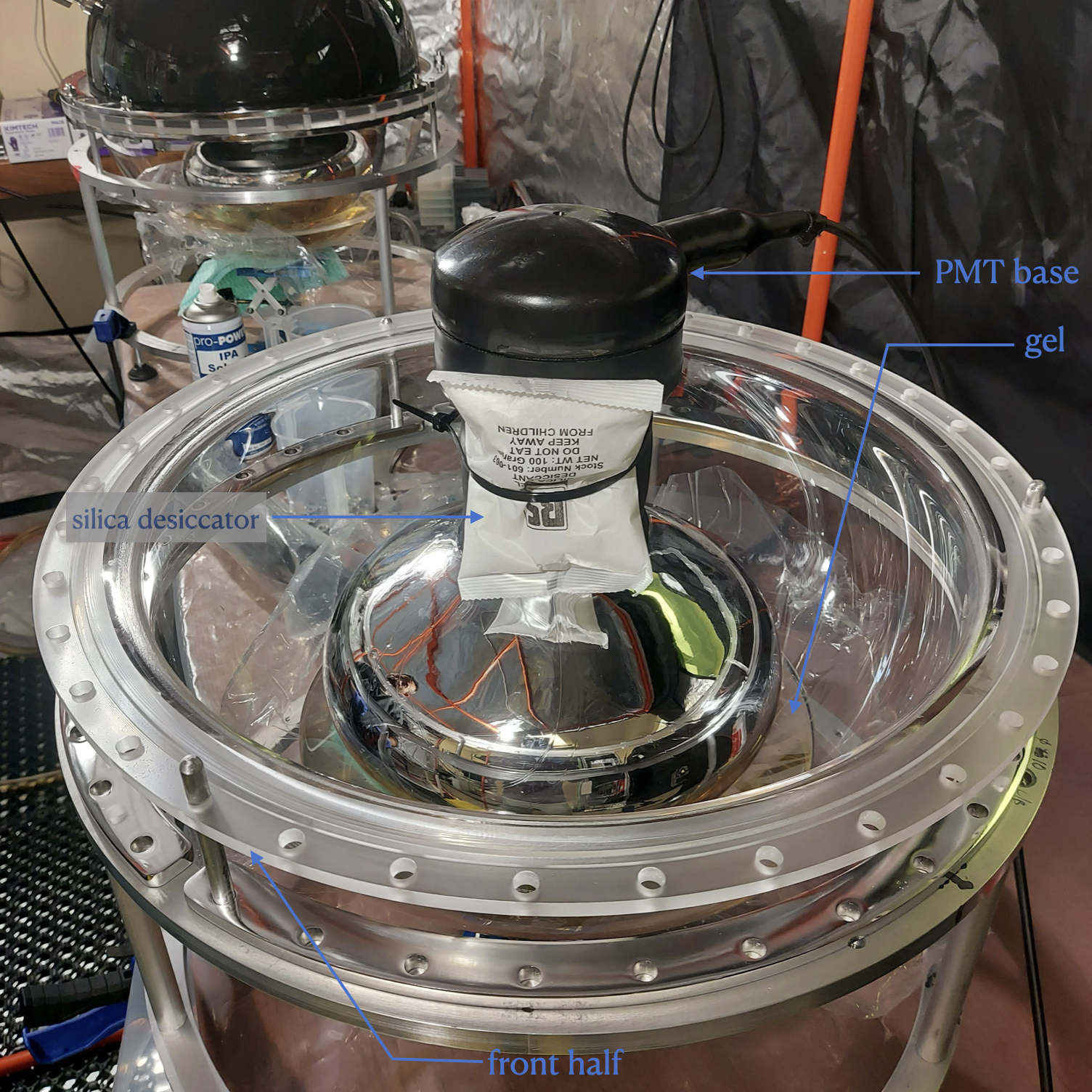}
\caption{\label{fig:12-degassing} Left: degassed RTV27905 gel in the front hemisphere, showing bubble-free optical contact medium. Right: PMT embedded in the cured gel after 24 hours, demonstrating mechanical stability without external support.}
\end{figure}

%\subsubsection{Preparing the optical gel}
%\label{subsubsec:OpticalGel}
The next step was to thoroughly mix the two components of the optical gel in equal amounts. This process inevitably introduced air bubbles into the mixture. To remove the bubbles, the mixed gel was poured into the front half and degassed under vacuum for 20 minutes (Figure~\ref{fig:12-degassing}). 

%\subsubsection{Curing the coupled system}
Once degassing was complete, the acrylic lid was removed. The PMT, secured in the holder, was lowered into the gel. The rig and holder were adjusted so that a gap of 4\,mm was maintained between the photocathode and the front hemisphere, allowing the gel to spread uniformly over the photocathode surface. This procedure was completed within the 30-minute work time of the mixed gel specified by the manufacturer. Once positioned, the PMT and holder were left for 24 hours to allow the gel to cure. The room temperature and humidity during curing was around $24 \pm 2~^\circ\mathrm{C}$ and $45 \pm 5\%$ respectively.

\begin{figure} %%%%%%%%%%%%%%% Fig %%%%%%%%%%%%%
\includegraphics[width=0.49\linewidth]{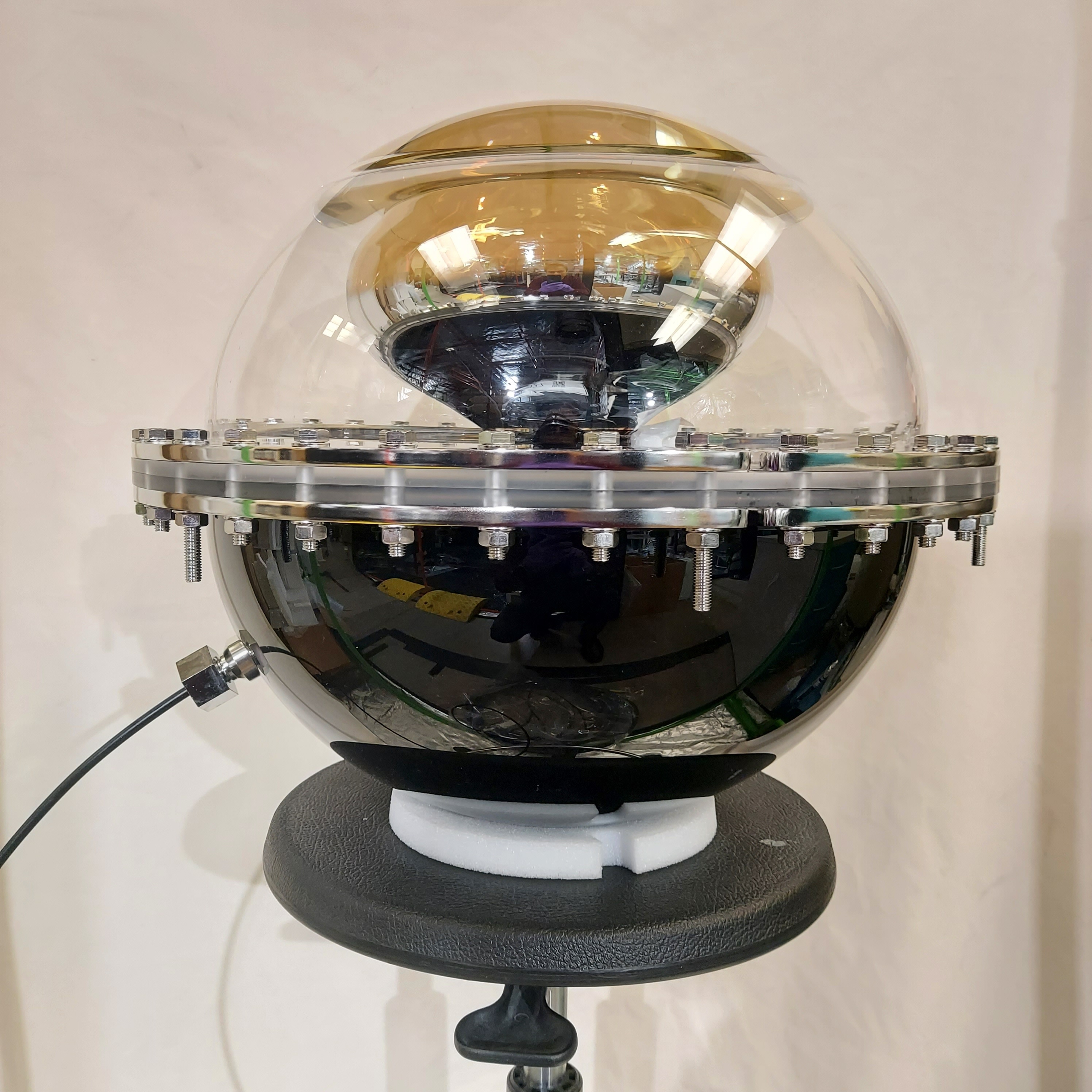}
\includegraphics[width=0.49\linewidth]{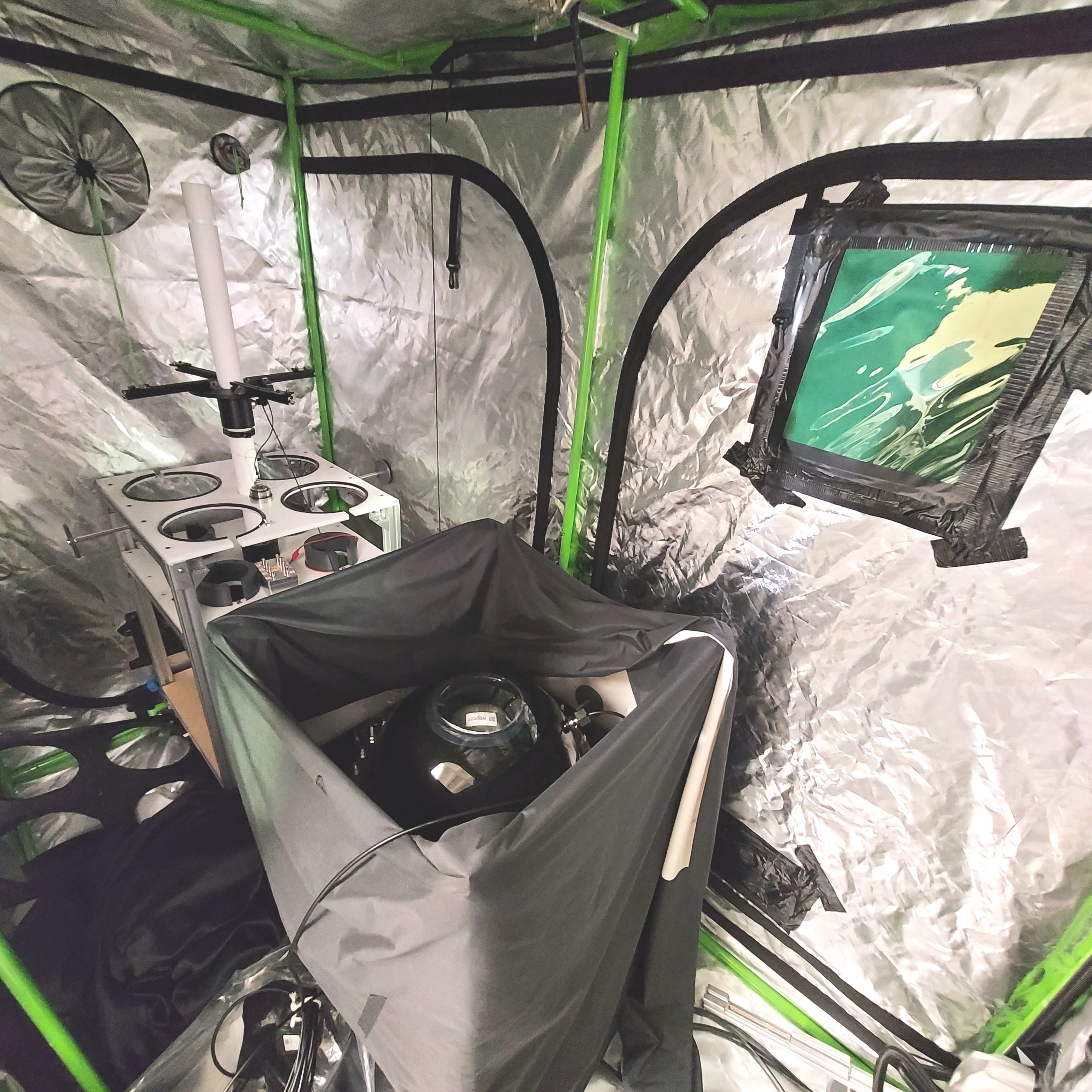}
\caption{\label{fig:14-complete-module} Left: fully assembled optical module with encapsulated PMT and penetrator system. Right: module placed inside a light-tight dark tent for electrical and optical quality assurance tests, including dark count and current draw measurements.}
\end{figure}

\subsection{Closure of the module halves}
After 24 hours, the PMT holder was removed, leaving the PMT securely embedded in the gel (Figure~\ref{fig:12-degassing}). The back half of the housing was then threaded through the HV cable and placed over the PMT. It was aligned so that the internal collar of the back half sat securely against the PMT base, providing mechanical support. A thin foam pad was inserted between the collar and the PMT base to fill any residual gap. At this stage, it was verified that the O-ring was properly seated, the housing was labelled inside with the PMT number, and the silica pouch was attached to the PMT base using a zip tie (Figure~\ref{fig:12-degassing}).

\subsection{Torquing}
Once the back half was placed on the front half, three additional washer plates were placed on the flange, with the alignment pins ensuring proper positioning. The two halves were coupled using six long bolts and thirty short bolts (Figure~\ref{fig:CAD-optical-module}). A predefined pattern was followed for the placement of the long bolts relative to a reference, as these secure the optical module to the PMT-support frame within the tank.

The bolts were tightened in stages using torque wrenches ($4\%$ tolerance): an initial pass at 4.5 Nm, followed by two passes at 6 Nm, using a cross pattern to ensure uniform tension across the flange. Finally, the outer cap nut of the penetrator was tightened to 12 Nm. A checklist was maintained to confirm that this procedure was followed. A fully assembled module is shown in Figure~\ref{fig:14-complete-module}.

\subsection{Quality assurance and shipping}
\label{subsec:QAQC}
After assembly, each module was immersed in a tank containing deionised water for at least 30 minutes to test the water tightness. Two modules required re-work at this stage due to incorrect torque procedures being followed (missed final pass on the cap nut). In addition to the standard immersion test, the selected modules were kept underwater for up to 2 weeks without any signs of water intrusion. As a further check, a module was cooled to 0$^{\circ}~\mathrm{C}$\ in a climate chamber and then immersed in the water tank without problems. 

After completion of the immersion test, the HV connector was installed to allow electrical testing. Each module was placed in a tent (Figure~\ref{fig:14-complete-module}) and the dark count and the current draw were measured at the nominal operating voltage. The values obtained were verified to match those obtained in previous tests~\cite{Akindele:2023ixz}. 

\begin{figure} %%%%%%%%%%%%%%%% Fig9 %%%%%%%%%%%%%%%%%%%%%%%%
\centering
\includegraphics[scale=0.55]{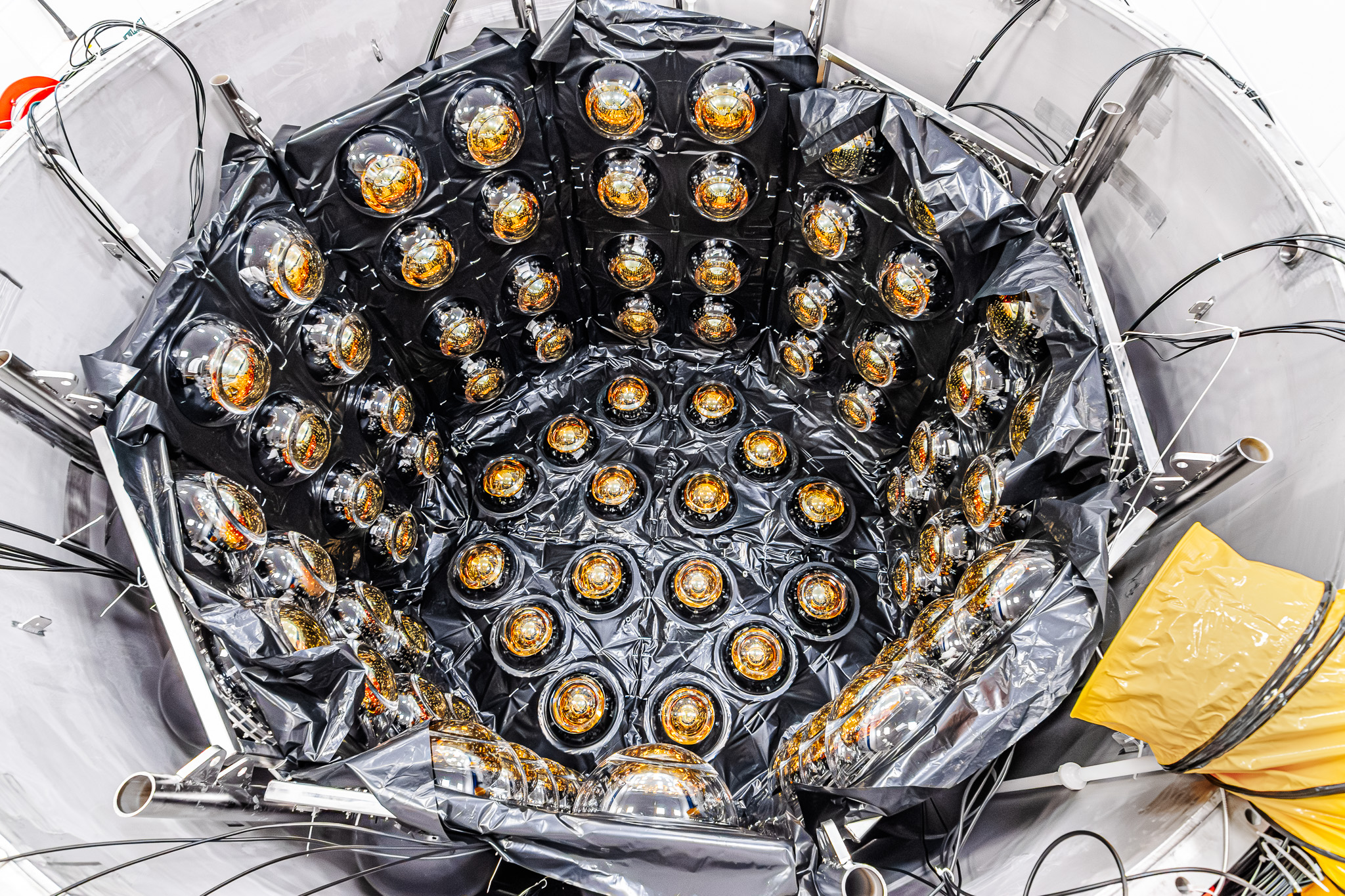}
\caption{Installation of encapsulated PMTs inside the BUTTON-30 detector tank at the STFC Boulby Laboratory. The photograph shows sixty-four radial and sixteen bottom-mounted modules positioned within the support structure, ready for operation with different fill media.}
\label{fig:16-PMTs-inTank}
\end{figure}

After completion of the quality assurance process, each module was cleaned with deionised water and sealed in a plastic bag. For safe transportation, the module was secured within a cardboard box, cushioned on all sides with air-foam inserts. The box was then wrapped in a thin plastic film.  This procedure was essential to protect the module from dust contamination, particularly during transport inside the mine.  Before entering the Class-10000 cleanroom space of the Boulby laboratory, the outer bag was removed. Before installation, each module was removed from the inner bag and final functionality checks were made.  This step included a visual inspection, a bolt torque check, and a verification of the impedance of the base.  All modules passed these tests and were installed in the tank without issue (Figure~\ref{fig:16-PMTs-inTank}).  

%tank_w_PMT

As a further check, a spare module was shipped to Brookhaven National Laboratory (BNL). This module was operated in $1\%$ WbLS for several weeks with no visible signs of corrosion, degradation or cracks.

\section{Summary}
Ninety-nine optical housings were designed and produced to encapsulate the PMTs of the BUTTON-30 detector. Novel encapsulation of the PMTs allows the detector to test new media, such as WbLS and gadolinium loadings. The production process was efficient, with clearly defined steps and robust quality assurance, resulting in $98\%$ of modules being successfully deployed on the first attempt. The PMTs are now installed in the detector tank at Boulby and are ready to collect data. The developed acrylic housing can withstanding a pressure of 3\,bar and are suitable for use in larger-volume detectors.
Future work will include long-term monitoring of acrylic stability in WbLS, as well as in-situ optical performance measurements once BUTTON-30 begins data-taking.

\noindent{}\textbf{Data Availability} Data supporting this work are available from the corresponding author upon reasonable request.

\section*{Acknowledgements}
The Boulby Underground Laboratory is funded and operated by the Science and Technologies Facilities Council (STFC) operating under United Kingdom Research and Innovation (UKRI). The laboratory is located in an active polyhalite mine operated by ICL Boulby. The authors thank both the STFC Boulby and ICL staff. Without their cooperation and support, the installation of the detector and optical modules would not have been possible.

We thank Doug Cowan and John Learned for their insightful and fruitful discussions on housing design. The housing design builds on the IceCube DOM concept and we express our gratitude to our colleagues at the University of Wisconsin (in particular, Sam Wolcott) for providing details of the IceCube DOM assembly procedure.

This work was supported in the UK by several STFC grants (ST/S006400/1, ST/S006419/1, ST/S006524/1, ST/T005319/1, ST/V002341/1, ST/Y003373/1, ST/Y003500/1, ST/Y003462/1, ST/Y003411/1) and by the Atomic Weapons Establishment (AWE), as contracted by the Ministry of Defence.
The work of Ben Richards is supported by the UKRI grant MR/Y034082/1.
The collaboration with ICLTech to develop the housings was supported by an STFC Accelerator grant awarded to G.~D. Smith. Testing a fully assembled module at BNL was facilitated by a NuSec travel award to D.~S. Bhattacharya and M. Needham. 

Work was performed under the auspices of the U.S. Department of Energy by Lawrence Livermore National Laboratory under contract DE-AC52-07NA27344, release number LLNL-JRNL-2013097. The work was supported in part by the U.S. Department of Energy National Nuclear Security Administration Office of Defence Nuclear Nonproliferation R\&D through the Consortium for Monitoring, Technology, and Verification under award number DE-NA0003920. The work carried out at Lawrence Berkeley National Laboratory was carried out under the auspices of the U.S. Department of Energy under the contract DE-AC02-05CH11231. The project was funded by the U.S. Department of Energy, National Nuclear Security Administration, Office of Defense Nuclear Nonproliferation  Research and Development (DNN R\&D). This material is based upon work supported by the U.S. Department of Energy, Office of Science, Office of High Energy Physics, under Award Number DE-SC0018974.

For the purpose of open access, the author has applied a Creative Commons Attribution (CC BY) licence to any Author Accepted Manuscript version arising from this submission.

%%===========================================================================================%%

\bibliography{sn-bibliography}% common bib file

\end{document}